\documentclass{appolb}
\usepackage{epsfig}

\begin{document}
\def\prl{{\em Phys. Reports }}
\def\prl{{\em Phys. Rev. Lett. }}
\def\prc{{\em Phys. Rev. {\bf C} }}
\def\prd{{\em Phys. Rev. {\bf D} }}
\def\jap{{\em J. Appl. Phys. }}
\def\ajp{{\em Am. J. Phys. }}
\def\nima{{\em Nucl. Instr. and Meth. Phys. {\bf A} }}
\def\npa{{\em Nucl. Phys. {\bf A}}}
\def\npb{{\em Nucl. Phys. {\bf B}}}
\def\epjc{{\em Eur. Phys. J. {\bf C}}}
\def\epja{{\em Eur. Phys. J. {\bf A}}}
\def\plb{{\em Phys. Lett. {\bf B}}}
\def\mpla{{\em Mod. Phys. Lett. {\bf A}}}
\def\pr{{\em Phys. Rep.}}
\def\zpc{{\em Z. Phys. {\bf C}}}
\def\zpa{{\em Z. Phys. {\bf A}}}
\def\app{{\em Acta Physica Polonica {\bf B}}}
\def\rnc{{\em Riv. Nuovo Cimento}}
\def\jpg{{\em J. Phys. G: Nucl. and Part. Phys.}}
\def\epjst{{\em Eur.Phys.J.ST}}
\def\epja{{\em Eur.Phys.J. {\bf A}}}
\def\njp{{\em New J. Phys.}}

\title{$\phi$ Meson As a Probe of QCD Equation of State}

\author{Raghunath Sahoo$^{1,2}$\footnote{Speaker},Tapan K. Nayak$^3$, 
Jan-e Alam$^3$, Sonia Kabana$^2$, Basanta K. Nandi$^4$, and Durga P. Mahapatra$^5$
\address{$^1$INFN, Sezione di Padova, 35131-Padova, Italy\\
$^2$SUBATECH, 4, Rue Alfred Kastler, BP 20722 - 44307 Nantes, France\\
$^3$Variable Energy Cyclotron Centre, 1/AF Bidhan Nagar, Kolkata 700064, India\\
$^4$Indian Institute of Technology, Powai, Mumbai 400076, India\\
$^5$Institute of Physics, Sachivalaya Marg, Bhubaneswar 751005, India}
}
\maketitle

\begin{abstract}
In this work, we extract the QCD Equation of State (EoS) using experimental 
results of the $\phi$ meson produced in nuclear collisions at AGS, SPS and 
RHIC energies. The data are confronted to simple thermodynamic expectations 
and lattice results. The experimental data indicate a first order phase 
transition, with a mixed phase stretching energy density between $\sim$ 
1 and 3.2 GeV/$fm^3$.
\end{abstract}
\section{Introduction}
High energy heavy-ion collisions provide an unique opportunity to test the Quantum 
Chromodynamics (QCD) prediction of a phase transition from hadronic matter 
to a deconfined thermalized state of quarks and gluons called Quark 
Gluon Plasma (QGP). The critical temperature for this
transition is $T_c\sim 170$ MeV \cite{karsch}. The order of such a phase transition
is still a matter of debate. Current lattice QCD calculations indicate that the 
order of the transition depends on the quark masses, as well as on the 
baryochemical potential ($\mu_B$). The magnitude of the baryochemical potential 
at the central rapidity region depends on the collision energy, it reduces with
the increase in beam energy. Therefore, by changing the beam energy it is possible
to study the change in the nature of QCD phase transition.

\noindent
~~~In a first order phase transition scenario, the pressure increases with 
increasing temperature, until the transition temperature $T_c$ is reached, 
then it remains constant during the mixed phase, and continues to increase 
after the end of the mixed phase. Related to this picture, L. Van Hove 
\cite{vanHove} suggested to identify the deconfinement transition in high 
energy $p\bar{p}$ collisions, looking at the variation of average transverse 
momentum ($\left<p_T\right>$) of hadrons as a function of the hadron 
multiplicity at midrapidity ($dN/dy$) and searching for an increase of 
$\left<p_T\right>$ followed by a  plateau-like behavior and again a subsequent 
increase. The $\left<p_T\right>$ is expected to reflect the thermal freeze-out 
temperature $(T_{th})$ and a flow component which can be related to 
the initial pressure, similarly the hadronic multiplicity reflects the entropy 
density of the system. While the purely thermal component of $\left<p_T\right>$ 
can not be related to the initial temperature which remains unmeasurable above 
$T_c$, however the flow component in the inverse slope can reflect the plateau 
of the pressure during mixed phase.

\noindent
~~~It has been observed that in central heavy ion collisions, the 
$\left<m_T\right>(=\sqrt{p_T^2+m^2})$ of pions, kaons and protons as a 
function of $dN_{ch}/dy$ 
shows a Van-Hove-like behaviour as explained above for a wide range of 
collision energies \cite{bm,marek}. Hydrodynamic calculations assuming a 
first order transition could reproduce these data \cite{bm,kodama}. In this 
work, we study for the first time the variation of the inverse slope, 
$T_{\textrm{eff}}$ extracted from the $p_T$ distribution of the $\phi$-meson 
as a function of the initial energy density, $\epsilon_{Bj}$  evaluated 
within the framework of Bjorken's hydrodynamical model~\cite{bjorken} and  
$\sqrt{s_{NN}}$ for energies spanning from AGS, SPS to RHIC \cite{raghu1}. 
This is an important 
new feature of such studies, because it connects the inverse slope with a 
parameter characterizing the initial state of the collision build up after
$\sim$ 1 fm/c and which reflects at the same time the collision energy, 
the stopping and the impact parameter of the collision.  For example, at a 
given $\sqrt{s_{NN}}$, different energy densities could be achieved by 
changing the colliding nuclei species or the impact parameter. 
$\epsilon_{Bj}$ can be directly compared to the critical energy density $\epsilon_c$ 
obtained in lattice QCD calculations- $\epsilon_c\sim 1$ GeV/fm$^3$. In the 
following the $\phi$ data will be analysed and confronted to simple 
thermodynamic expectations which relate to the Van Hove signature and to 
lattice QCD predictions.  Due to its 
$s\bar{s}$ valence quark content, the $\phi$ meson is of special interest 
\cite{bedanga_sqm2008} to study strangeness enhancement, which is a potential 
probe of QGP formation. Because of its small hadronic rescattering cross 
section of $\sigma$($\phi$N)=10 mb, it decouples earlier than other hadrons. 
Furthermore, due to its life time of $\sim$ 45 fm/c, its main decay product 
$(K^+K^-)$ suffer  less  rescattering. Experimental results from Au+Au 
collisions at RHIC energies indicate  that $\phi$ has a higher thermal 
freeze out temperature as compared to pions, kaons and protons. In particular, 
their thermal freeze out temperature is within errors compatible with the 
chemical freeze out temperature of hadrons and the critical temperature. 
Therefore $\phi$ and its flow phenomena are particularly interesting 
probes for studying the EoS and the nature of the phase transition.

\section{$\phi$ as a probe of the order of the phase transition}

\begin{figure}
\begin{center}
\resizebox{0.8\columnwidth}{!}{%
\includegraphics{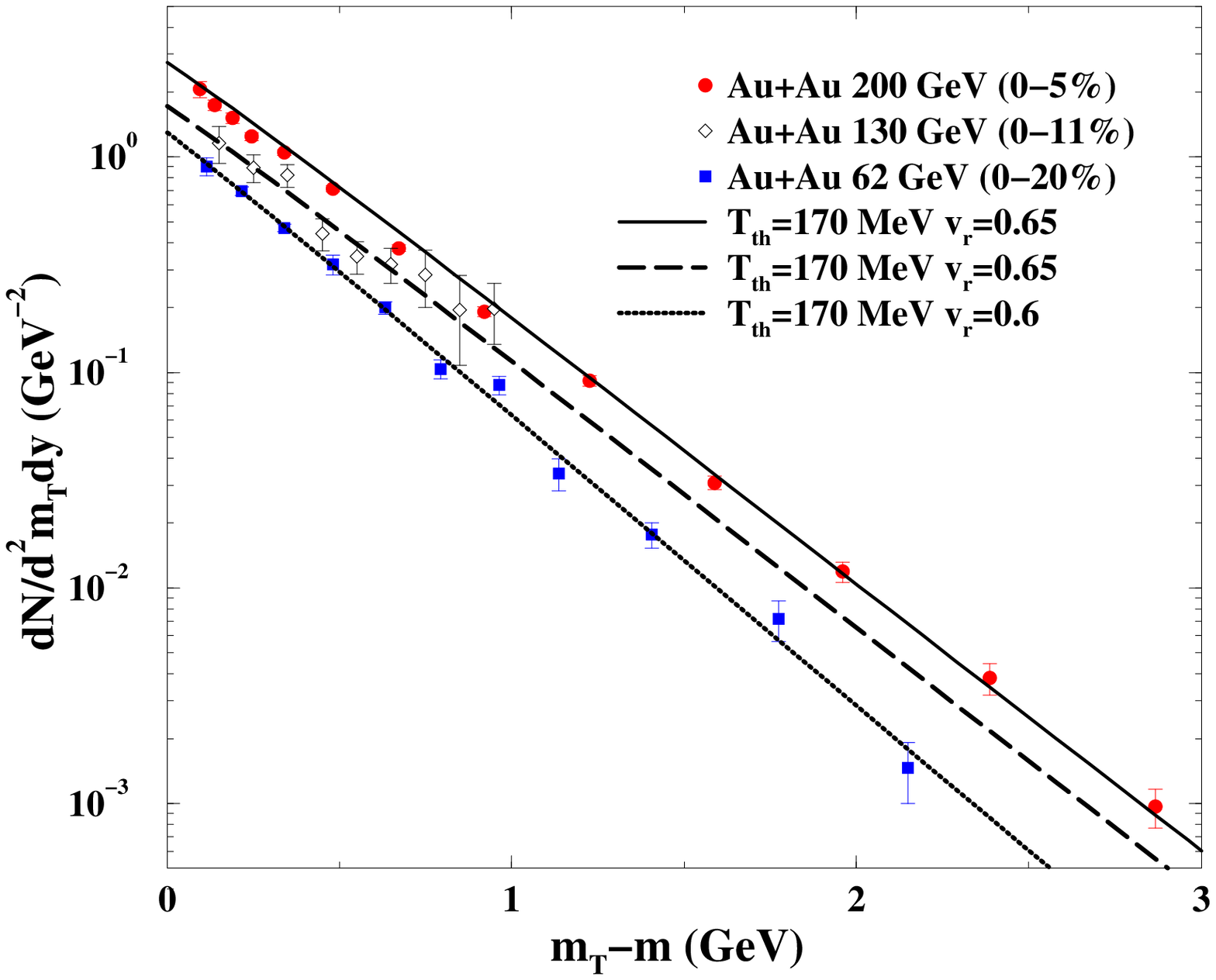}
\includegraphics{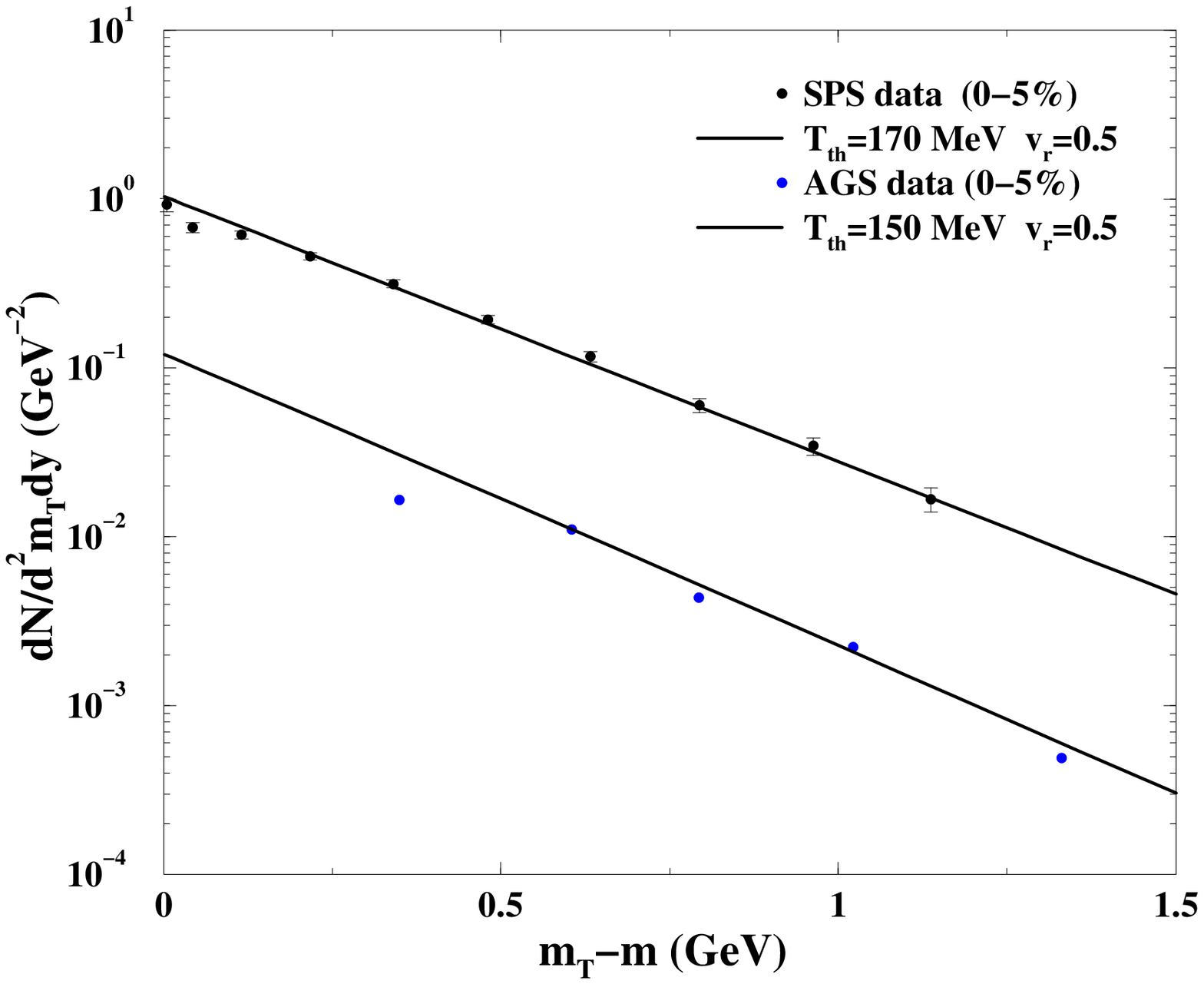}}
\caption{Transverse mass distribution of $\phi$ as a function of its
kinetic energy at mid-rapidity for different colliding energies measured
by the STAR collaboration at RHIC (left) and for SPS and AGS energies (right). 
The theoretical curves are obtained within the framework of blast wave method.} 
\label{starphi}
\end{center}
\end{figure}

\noindent
The $p_T$ spectra of $\phi$ measured in heavy ion collisions at various
collision energies~\cite{rhicPhi,spsPhi,agsPhi} have been reproduced within 
the ambit of the blastwave method~\cite{blastW}, results for AGS, SPS and 
RHIC are shown in figure ~\ref{starphi}. The values of the parameters {\it i.e.} 
the radial flow velocity, $v_r$ and the freeze-out temperature, $T_{th}$ of 
the blastwave method are displayed in table I (see also~\cite{nu}). 
It has been found that $T_{th}\sim$ 170 MeV and $v_r\sim$ 0.6 and are independent 
of the center of mass energies beyond a thershold. $T_{th}$ being close to $T_c$ 
indicates that $\phi$ meson 
freezes out near the phase boundary and could be used to extract the properties of 
QCD matter near the transition point. Moreover the values of $v_r$
indicate that  QGP has undergone substantial radial flow.
The values of the `true' temperature and the flow velocity of the system 
at the $\phi$ freeze-out surface for different colliding energies 
can be used to get the effective inverse slope parameter
$T_{eff} = T_{th} + \frac{1}{2}m\left<v_r\right>^2$. 

\begin{table}
\begin{center}
\begin{tabular}{lcccr}
\hline\hline
$\sqrt{s_{NN}}$(GeV)&$T_{th}$(MeV)&$v_r$\\
\hline
4.87&150&0.5\\
17.3&170&0.5\\
62.4&170&0.6\\
130&170&0.65\\
200&170&0.65\\
\hline\hline
\end{tabular}
\caption{Freeze-out temperature, $T_{th}$ and the radial flow velovity, $v_r$ for 
$\phi$-meson for various centre of mass energies, $\sqrt{s_{NN}}$ extracted from the
experimental data within the framework of blast wave method.}
\end{center}
\end{table}

\begin{figure}
\begin{center}
\resizebox{0.8\columnwidth}{!}{%
\includegraphics{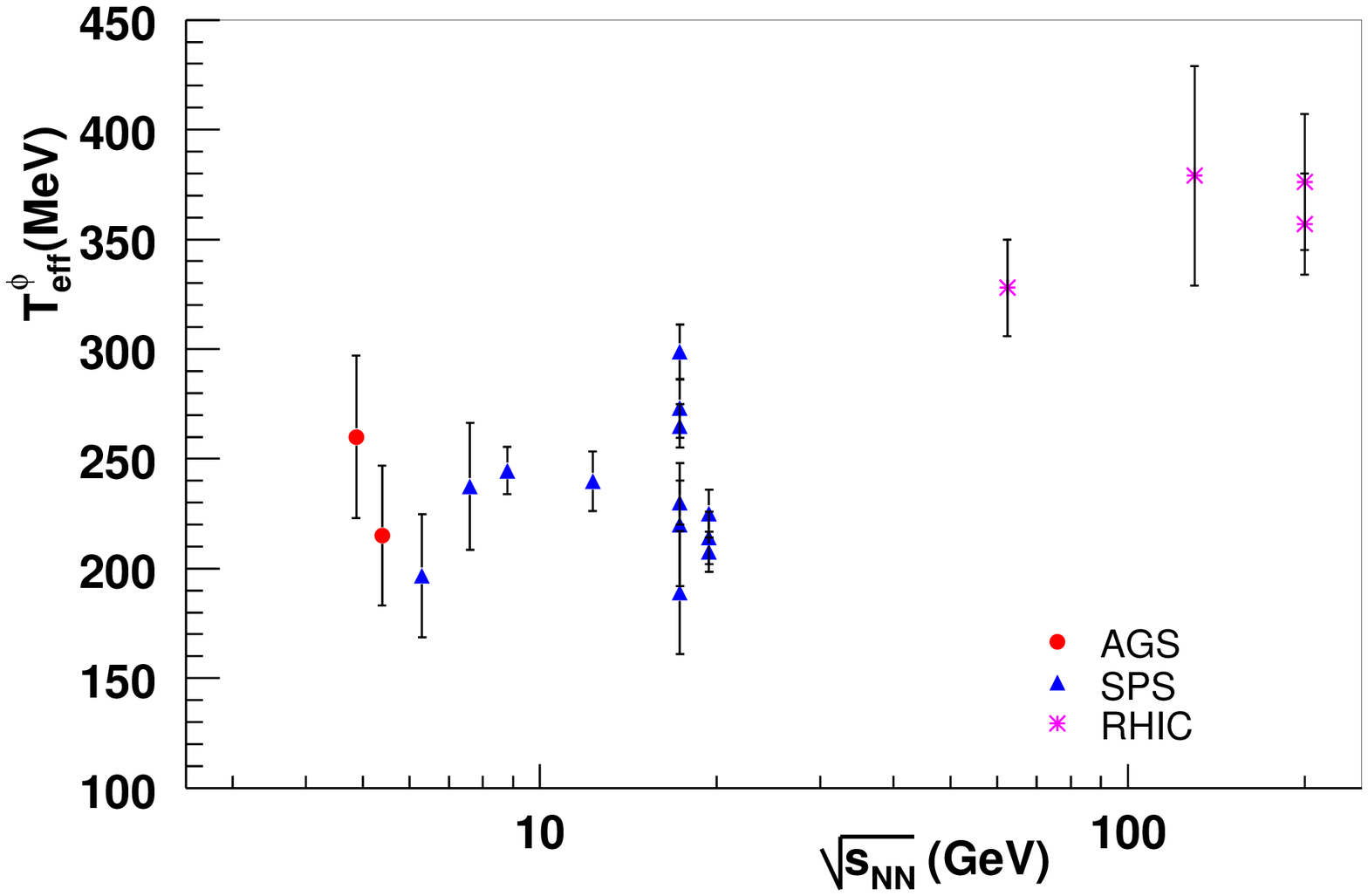}
\includegraphics{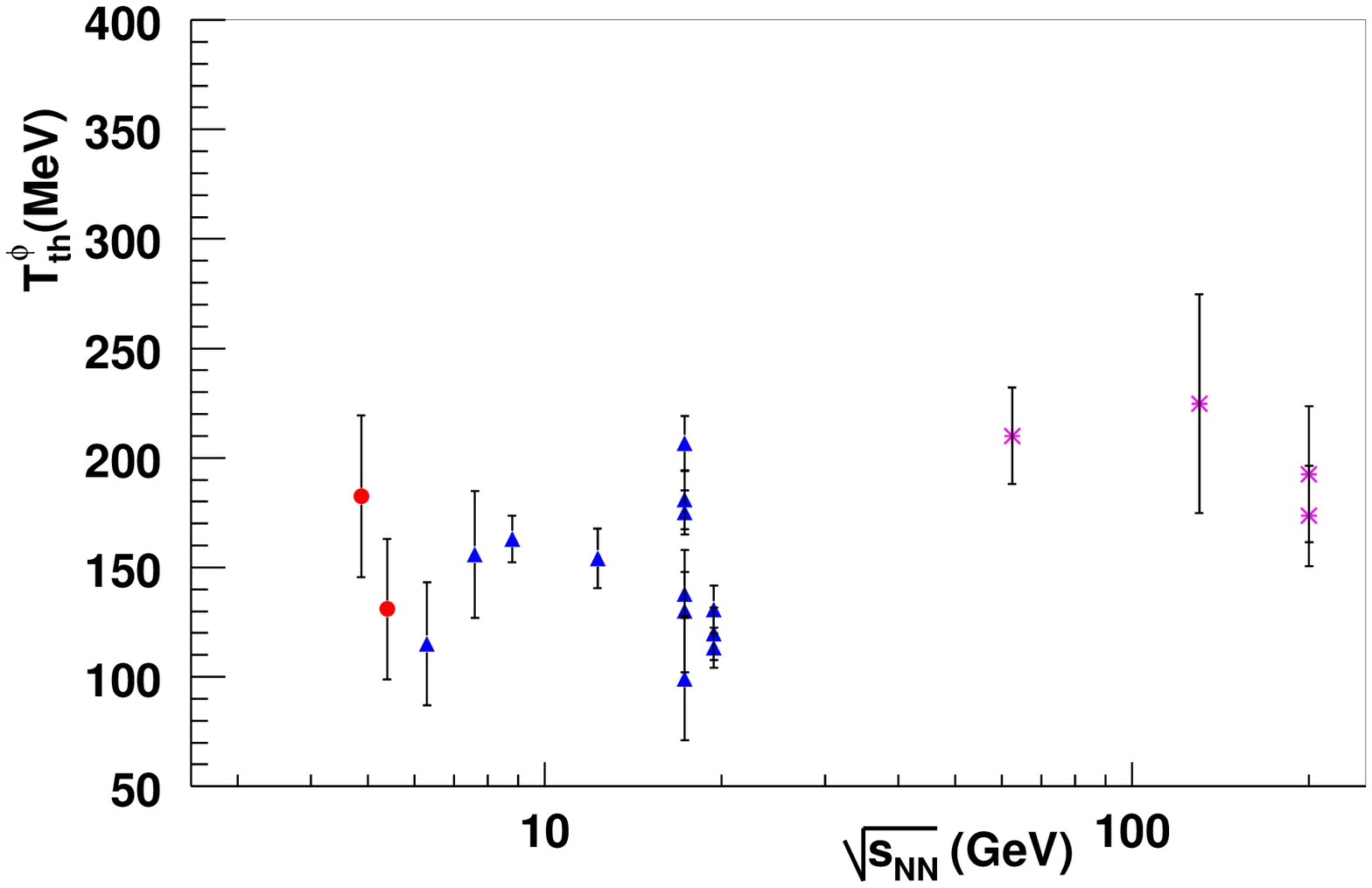}}
\caption{$T_{\textrm{eff}}^{\phi}$ (left ) and $T_{\textrm{th}}^{\phi}$ (right) 
as a function of $\sqrt{s_{NN}}$ from AGS-SPS to RHIC.}
\label{withCoM}
\end{center}
\end{figure}

\noindent
~~~In figure \ref{withCoM} the effective temperature of $\phi$, 
$T_{\textrm{eff}}^{\phi}$ (left) and the thermal component of the inverse 
slope $T_{\textrm{th}}^{\phi}$ (right) are shown as a function of the collision 
energy $\sqrt{s_{NN}}$. We use the central collision data in almost the same 
$p_T$ range, at mid-rapidity \cite{rhicPhi,spsPhi,agsPhi,floris,na38,na50,na49}. 
It is observed 
that from AGS to SPS energies these observables remain almost unchanged, showing 
a plateau-like structure. Going from SPS to RHIC energies, the inverse slope 
$T_{\textrm{eff}}^{\phi}$ exhibits a sudden jump while an increase is still 
observed in the  $T_{\textrm{th}}^{\phi}$ component. This may be due to an 
imperfect transverse flow component subtraction at RHIC or other effects. It is 
observed that the $T_{\textrm{th}}^{\phi}$ is reaching at RHIC  values compatible 
within errors with $T_c$. The observed plateau of $T_{\textrm{eff}}^{\phi}$ is a 
signature of a coexisting phase of quarks, gluons and hadrons for a first order 
phase transition, during which the initial pressure remains constant. The 
subsequent increase of $T_{\textrm{eff}}^{\phi}$ with $\sqrt{s_{NN}}$ at top 
RHIC energies indicates the end of the mixed phase and the entering into a pure 
QGP phase.

\begin{figure}
\begin{center}
\resizebox{0.8\columnwidth}{!}{%
\includegraphics{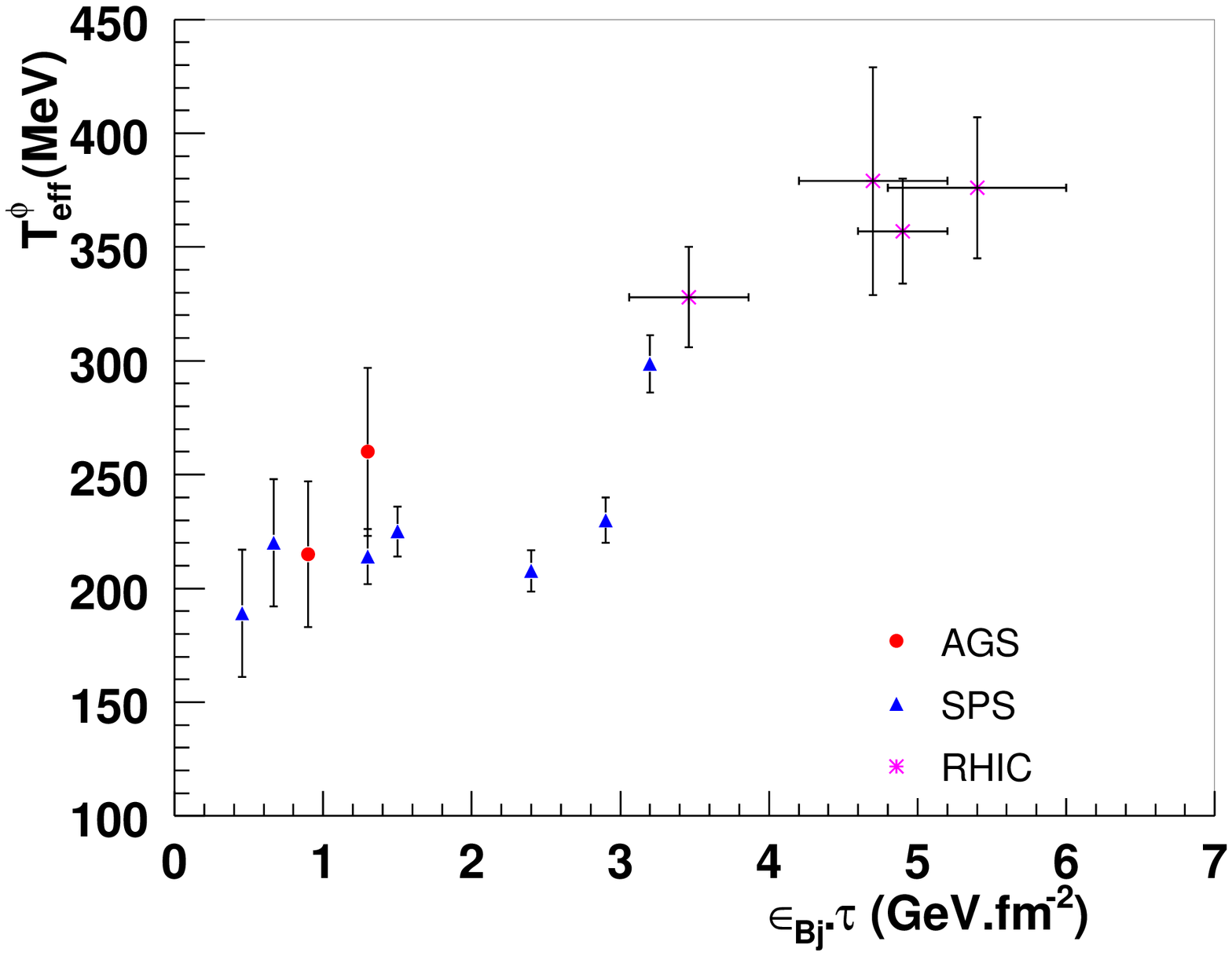}
\includegraphics{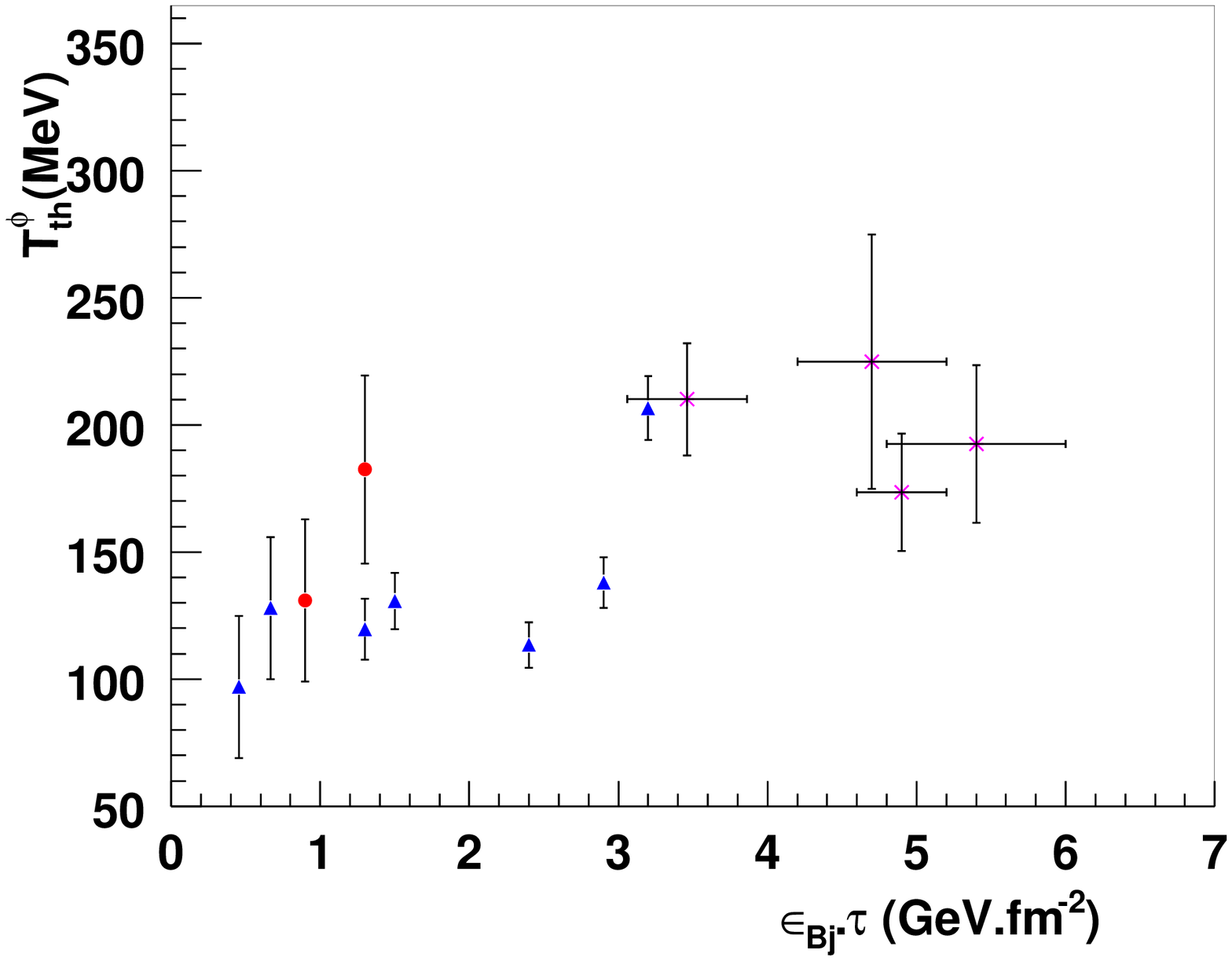}}
\caption{$T_{\textrm{eff}}^{\phi}$ and $T_{\textrm{th}}^{\phi}$ as 
a function of $\epsilon_{\textrm{Bj}}.\tau$ from AGS-SPS to RHIC.}
\label{withEDensity}
\end{center}
\end{figure}

\noindent
~~~Now we study the dependence of the inverse slope and its thermal component
on  the initial energy density, $\epsilon_{Bj}$, estimated as
$ \epsilon_{Bj} = \left<\frac{dE_T}{dy}\right>\frac{1}{\tau\pi R^2} = 
\left<\frac{dN}{dy}\right>\left<m_T\right> \frac{1}{\tau\pi R^2}$
where $R = R_0 A^{1/3}$ and $A\sim N_{part}/2$. From the experimental 
measurements of $dE_T/dy$ and $dN/dy$ with $\left<m_T\right>$,
the observable $\epsilon_{Bj}$ could be estimated for all centralities 
and center of mass energies at mid-rapidity.

\noindent
~~~In figure \ref{withEDensity} $T_{\textrm{eff}}^{\phi}$ (left) and 
$T_{\textrm{th}}^{\phi}$ (right) are shown as a function of $\epsilon_{Bj}.\tau$. 
Note that the formation time, $\tau$ is model-dependent and in general, in 
subsequent discussions we also assume $\tau \sim 1$ fm/c. The above result 
reflects the properties of equation of state. We observe a plateau in 
$T_{\textrm{eff}}^{\phi}$ stretching between $\epsilon_{Bj} ~\sim$ 1 and 
3.2 GeV/$fm^3$, and increasing suddently above 3.2 GeV/$fm^3$. This behaviour 
as already discussed, suggests a first order transition, however from figure 
\ref{withEDensity} 
we can now infer that the mixed phase is stretching between 1 and 3.2 GeV/$fm^3$.
The use of the $\epsilon_{Bj}$ scale, allows us to establish
here for the first time the $\epsilon_{Bj}$ range of the mixed phase.
More data on the $\phi$ at  $\epsilon_{Bj}$ below 1 GeV/$fm^3$ are needed
to establish the increase of $T_{\textrm{eff}}^{\phi}$ up to 1 GeV/$fm^3$, as 
seen in other hadrons. The increase of the inverse slope at RHIC energies 
again indicate a pure phase of QGP, as is expected from a first-order phase 
transition. The thermal component $T_{\textrm{th}}^{\phi}$ shows also a plateau 
while a smaller increase is still observed at 3.2 GeV/$fm^3$. The step-like 
behavior in the excitation function of $<m_T>$ has also been explained by taking 
a first-order phase transition with a large latent heat or if the EoS is 
effectively softened due to non-equilibrium effects in the hadronic transport 
calculations \cite{petersen}. 

\noindent
~~~ The flow component of the inverse slope of the $\phi$ reflecting the 
initial pressure, is of interest to look directly observables linked to 
this initial pressure like the  transverse flow velocity, $v_r$ for all 
hadrons and the elliptic flow $v_2$ as a function of collision energy.
The transverse flow  velocity $v_r$  shows exactly the same characteristic 
behaviour as the $T_{\textrm{eff}}^{\phi}$ as a function of collision energy, 
namely increase, a plateau and subsequent increase at RHIC \cite{nu}.
A similar pattern is suggested for the elliptic flow $v_2$ as a function of 
collision energy \cite{lokesh_sqm2008}. The observation of a mixed phase goes 
inline with the fact that excitation function of various observables show 
anomalous behavior or saturation effects starting lower SPS energies 
\cite{raghu,alt}. This could be related to the onset of deconfinement 
corresponding to this energy regime. In addition, the fact that microscopic 
transport models, based on hadronic degrees of freedom failed to reproduce 
the observed behavior of kaon inverse slope \cite{el,wagner}, also indirectly 
confirms the observation of a deconfinement transition.

\section{Comparison of $\phi$ data with lattice predictions}

\begin{figure}
\begin{center}
\resizebox{0.9\columnwidth}{!}{%
\includegraphics{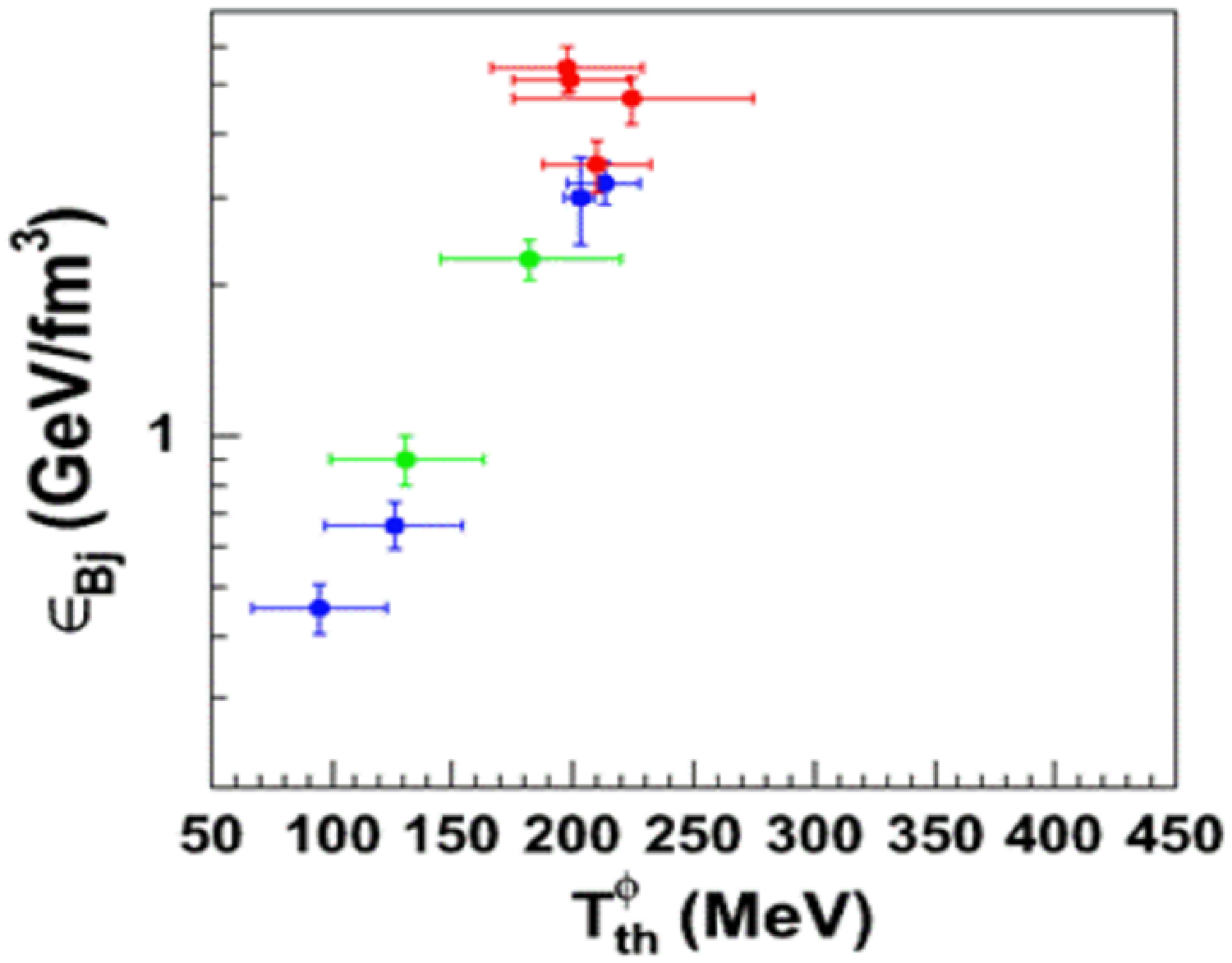}
\includegraphics{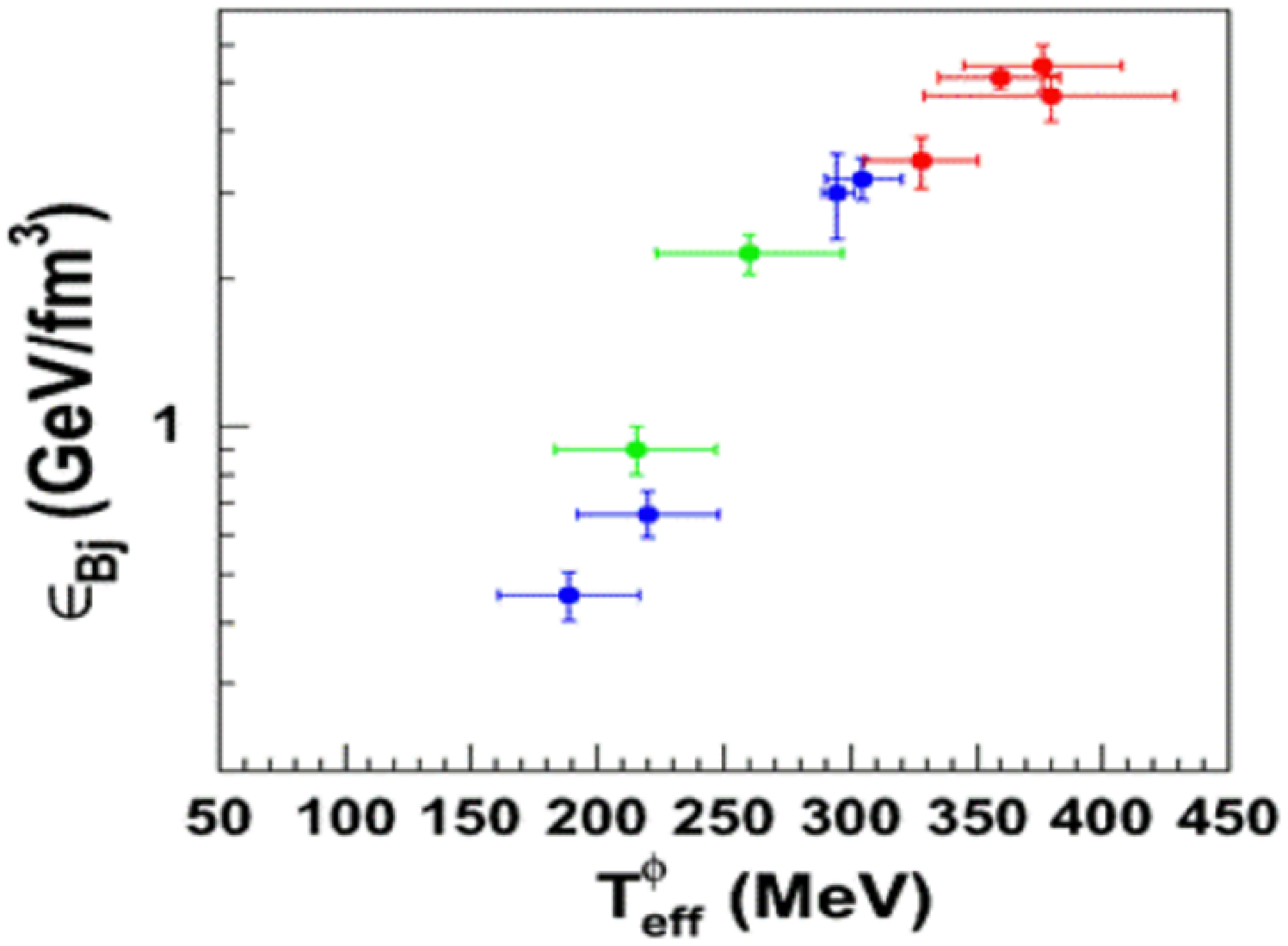}}
\caption{Energy density as a function of $T_{th}$ (left) and $T_{eff}$ 
(right) for the $\phi$ meson.}
\label{withLattice}
\end{center}
\end{figure}

\noindent
~~~In the following, we compare experimental data with lattice QCD predictions 
on energy density as a function of temperature as shown in figure 8 of \cite{lQCD}.
In figure \ref{withLattice}, the lattice prediction for the energy density  
as a function of the temperature is compared to $\epsilon_{Bj}$ as a function 
of $T^{\phi}_{th}$ (left) and $T^\phi_{eff}$ (right) extracted from the slopes 
of $\phi$ spectra. First we discuss the left plot. The lattice prediction is for 
zero baryochemical potential therefore corresponds to the cross over region. 
The figure in the left is at non-zero $\mu_B$ and shows the variation of the 
initial  energy density with $T^\phi_{th}$ temperature, which is measured at a 
later time than the energy density, namely at the thermal freeze out of the 
$\phi$. That is, the two 
variables in the data are a measure of the system at different times, while 
the lattice estimate is independent of time. This temperature is expected to be 
always below $T_c$ and does not reflect the initial $T$, which may exceed $T_c$ 
depending on the collision energy. This plot can be directly compared to figure 
11 of \cite{strangeborder}, where the energy density from data has been studied 
as a function of T, while both x and y axis were estimated at the same time, 
namely at the hadronic chemical freeze out time and at $\mu_{B} =0$.
It is seen that the energy density increases approaching the $T_c$ from below, 
from which both the $T_c$ and critical exponents can be extracted 
\cite{strangeborder}.

\noindent
~~~If the transition occurs at $T_c \sim 170$ MeV as expected, the temperature 
of $\phi$ at the thermal freeze out should saturate below and near $T_c$, for
all values of the initial density (up to infinity), as seen in figure 
\ref{withLattice} (left panel). We observe a saturation in the value of 
$T^\phi_{th}$ near $T_c$, because $T^{\phi}_{th}\leq T_c$  as mentioned earlier. 
To measure temperatures above $T_c$ photons and dileptons are useful probes.  

\noindent
~~~We now discuss figure \ref{withLattice} (right panel), where the initial 
Bjorken energy density is shown as a function of the effective temperature  
of the  $\phi$ for collisions from AGS, SPS to RHIC. Now both the x and y axis 
are reflecting parameters at initial times therefore this plot is more 
appropriate to be compared to the lattice results. The effective temperature  
of the  $\phi$ here is a sum of a thermal freeze out temperature, which is 
expected to be always below $T_c$, and a non-thermal component due to transverse 
flow, which relates to the initial pressure and reflects the initial conditions 
above $T_c$. Therefore the variables in the two plots compared here are not 
exactly the same but they are correlated. Further analysis is needed to compare 
the data  to lattice e.g. using exactly the same variables in both estimates.
A study can be also done involving the elliptic flow $v_2$ and the transverse
flow velocity  $v_r$ as a function of the initial energy density.
Also a hydrodynamic calculation of the discussed variables is of interest.

\section{Summary}
In summary,  a first analysis of experimental data on the inverse slope 
parameter of $\phi$ mesons as a function of $\sqrt{s_{NN}}$ \& energy density 
from AGS, SPS to RHIC, and their comparison to simple thermodynamic expectations, 
suggest a first order phase transition as predicted by lattice QCD at non-zero 
baryochemical potentials. The mixed phase of quarks, gluons and hadrons, is 
found to stretch between $\epsilon_{Bj}$ of $\sim$ 1 and 3.2 GeV/$fm^3$. The 
latent heat density in a first order phase transition is $\epsilon_Q(T_c)-
\epsilon_H(T_c)=4B$, where $\epsilon_Q (\epsilon_H)$ is the energy density 
of QGP (hadrons) at $T_c$ and $B$ is bag constant. With $\epsilon_Q(T_c)\sim 3.2$ 
GeV/fm$^3$ and $\epsilon_H(T_c)\sim 1$ GeV/fm$^3$, we get a reasonable value 
for $B^{1/4}\sim 250$ MeV. If we ignore $B$, then $\epsilon_Q/\epsilon_H 
\sim g_Q/g_H\sim 3.2$ also a reasonable number comparable to results from 
lattice QCD \cite{karsch}, here $g_Q (g_H)$ is statistical degeneracy of QGP 
(hadrons). The 
plateau and subsequent increase of the inverse slope parameter of $\phi$
 above $\epsilon_{Bj}$  3.2 GeV/$fm^3$ is in agreement with data from pions, 
kaons and protons, and is as well observed in the transverse flow velocity 
$v_r$ reflecting the behaviour of the initial pressure. A first attempt 
to compare data to lattice QCD predictions is made. 

\noindent
~~~The above results, while supporting the order of the transition predicted 
by lattice at large $\mu_B$ up to $\epsilon_{Bj}\sim 3.2$  GeV/fm$^3$,
they  do not exclude a change of order of the transition at smaller baryochemical 
pontials. Which points to further work, towards  mapping out the order of the 
QCD phase transition as a function of energy and baryochemical potential.

\section*{Acknowledgement}
We thank Drs. M. Floris, V. Friese, D. Jouan, A. De Falco for providing SPS data.


\begin{thebibliography}{90}
\bibitem{karsch} F. Karsch, \npa{\bf 698}, 199c (2002).
\bibitem{vanHove} L. Van Hove, \plb{\bf 118}, 138 (1982).
\bibitem{bm} B. Mohanty {\it et al.}, \prc{\bf 68}, 021901 (2003).
\bibitem{marek} M.I. Gorenstein, M. Gazdzicki and K.A. Bugaev, \plb{\bf 567}, 175 (2003).
\bibitem{kodama} Y. Hama {\it et al}, \app{\bf 35}, 179 (2004).
\bibitem{bjorken} J.D. Bjorken, \prd{\bf 27} 140 (1983).
\bibitem{raghu1} Raghunath Sahoo {\it et al.}, \jpg{\bf 36}, 064071 (2009), Preprint: {\it 0901.3254}[nucl-ex].
\bibitem{bedanga_sqm2008} B. Mohanty and N. Xu, {\it Preprint} 0901.0313
\bibitem{rhicPhi} B.I. Abelev {\it et al.} (STAR Collaboration), {\it Preprint} 
  0809.4737, \\S.S. Adler {\it et al.} (PHENIX Collaboration), \prc{\bf 72}, 014903 (2005).
\bibitem{spsPhi} M.C. Abreu {\it et al.} (NA38 Collaboration), \epjc{\bf 44}, 375 (2005). 
\bibitem{agsPhi} Y. Akiba {\it et al.} (E-802 Collaboration), \prl{\bf 76}, 2021 (1996),\\
  B.B. Back {\it et al.} (E917 Collaboration), \prc{\bf 69}, 054901 (2004).
 \bibitem{floris} M. Floris {\it et al.} (NA60 Collaboration), \epjc{\bf 49}, 255 (2007).
\bibitem{na38} M.C. Abreu  {\it et al.} (NA38 Collaboration), \plb{\bf 368}, 239 (1996).
\bibitem{na50} B. Alessandro {\it et al.} (NA50 Collaboration), \plb{\bf 555}, 147 (2003).
\bibitem{na49} S.V. Afanasiev {\it et al.} (NA50 Collaboration), \plb{\bf 491}, 59 (2000).
  C. Alt {\it et al.} (NA49 Collaboration), \prl{\bf 94} 052301 (2005), \prl{\bf 78}, 
044907 (2008). 
\bibitem{blastW} E. Schnedermann, J. Sollfrank  and U. Heinz, \prc{\bf 48}, 2462 (1993). 
\bibitem{petersen} H. Petersen, J. Steinheimer, M. Bleicher, H. St$\ddot{o}$cker, 
\jpg{\bf 36}, 055104 (2009).
\bibitem{nu} N. Xu, \npa {\bf 751}, 109c (2005).
\bibitem{lokesh_sqm2008} L. Kumar, (for STAR collaboration) {\it Preprint} 0907.1943
\bibitem{raghu} J. Cleymans, R. Sahoo, D.P. Mahapatra, D.K. Srivastava and S. Wheaton,  
	\plb{\bf 660}, 172 (2008), \epjst{\bf 155}, 13 (2008) and \jpg{\bf 35}, 
	104147 (2008).
\bibitem{alt} C. Alt {\it et al.} (NA49 Collaboration), \prc{\bf 77}, 024903 (2008).
\bibitem{el} E.L. Bratkovskaya {\it et al.}, \prc{\bf 69}, 054907 (2004);
  E.L. Bratkovskaya {\it et al.}, \prl{\bf 92}, 032302 (2004).
\bibitem{wagner} M. Wagner, A.B. Larionov, and U. Mosel, \prc{\bf 71}, 034901 (2005).
\bibitem{lQCD} D.E. Miller, \pr{\bf 443}, 55 (2007).
\bibitem{strangeborder} S. Kabana, \epjc {\bf 21}, 545 (2001);
 S. Kabana, P. Minkowski, \njp {\bf 3}, 4 (2001).

\end{thebibliography}
\end{document}